%% file: paper.tex
\documentclass{article}
\usepackage{dcase2022_techrep,amsmath,graphicx,url,times,booktabs, tabularx}
\usepackage{amsmath, amssymb}
\usepackage{multirow}
\usepackage{tabularx}

\usepackage{mathtools}
\usepackage{xspace}

\makeatletter
\DeclareRobustCommand\onedot{\futurelet\@let@token\@onedot}
\def\@onedot{\ifx\@let@token.\else.\null\fi\xspace}

\def\etal{\emph{et al}\onedot}
\makeatother

\title{VIFS: An End-to-End Variational Inference for Foley Sound Synthesis}

\name{Junhyeok Lee$^{1\ast}$\thanks{$\ast$ Equal contribution},
      Hyeonuk Nam$^{2\ast}$,
      Yong-Hwa Park$^{2}$ 
      }
\address{$^1$ maum.ai Inc., Republic of Korea,\\          
        $^2$ Korea Advanced Institute of Science and Technology, Republic of Korea,\\
         jun3518@icloud.com, \{frednam, yhpark\}@kaist.ac.kr\\
 }

\begin{document}

\ninept
\maketitle

\begin{sloppy}

\begin{abstract}
\input{sections/0_abstract}
\end{abstract}


\section{Introduction}
\input{sections/1_introduction}

\section{VIFS}
\input{sections/2_VIFS}

\section{Experiments}
\input{sections/3_experiments}

\section{Results and discussions}
\input{sections/4_results_and_discussion}

\section{Conclusion}
\input{sections/5_conclusion}

\bibliographystyle{IEEEtran}
\bibliography{paper}
\end{sloppy}
\end{document}

%% file: sections/0_abstract.tex
The goal of DCASE 2023 Challenge Task 7 is to generate various sound clips for Foley sound synthesis (FSS) by ``category-to-sound" approach. ``Category" is expressed by a single index while corresponding ``sound" covers diverse and different sound examples. To generate diverse sounds for a given category, we adopt VITS, a text-to-speech (TTS) model with variational inference. In addition, we apply various techniques from speech synthesis including PhaseAug and Avocodo. Different from TTS models which generate short pronunciation from phonemes and speaker identity, the category-to-sound problem requires generating diverse sounds just from a category index. To compensate for the difference while maintaining consistency within each audio clip, we heavily modified the prior encoder to enhance consistency with posterior latent variables. This introduced additional Gaussian on the prior encoder which promotes variance within the category. With these modifications, we propose VIFS, variational inference for end-to-end Foley sound synthesis, which generates diverse high-quality sounds.

%% file: sections/1_introduction.tex

Foley sound synthesis (FSS) involves the generation of various sound effects for movies.
While FSS could be implemented by text-to-audio to create detailed text-conditioned sound effects \cite{audiogen, audioldm}, the focus of the DCASE 2023 Challenge Task 7 is to begin the challenge with a simpler task, ``category-to-sound", which aims to generate sounds based on a simple category (or class) index \cite{fss, fss_proposal}.
The objective is to generate a wide range of sound examples based on a given category, where each category is represented by a single index.
It is important to note that within each category of sound events, there exist diverse acoustic characteristics due to the various entities capable of producing those sound events \cite{dcase2021t4, filtaug}. For instance, consider the sound of a dog barking. There are dogs of different species, each has different sizes and each is grown up in different environments. Such diverse entities of dogs result in significant variations in their barking sounds. As a result, the category-to-sound problem presents a unique challenge in generating diverse and distinct sounds solely based on categorical information, while accounting for the inherent variations within each category.

To address the aforementioned challenge, we draw upon the advancements in text-to-speech (TTS) which shares similarities with the category-to-sound problem in that both aim to generate audio output.
Since TTS focuses on simulating the pronunciation of given input text, it provides valuable insights and methodologies that can be adapted to the category-to-sound synthesis.
Our approach is based on VITS \cite{vits}, which showed exceptional performance with an end-to-end framework.
VITS consists of conditional variational auto-encoder (cVAE) \cite{vae}, normalizing flow \cite{flow}, and generative adversarial network (GAN) \cite{gan}.
By incorporating VITS into the category-to-sound synthesis, we can effectively generate a wide range of sounds that align with the given categories using an end-to-end framework, while the baseline \cite{fss} requires multiple stages including auto-regressive model, VQ-VAE structure \cite{baseline_model} and vocoder \cite{hifigan}.

However, the category-to-sound problem presents a significant difference from the TTS approaches. While category-to-sound has to generate sound clips spanning the whole clip just from a category index, TTS focuses on generating short pronunciations based on phonemes and speaker identities. To consider the difference between TTS and category-to-sound, we made heavy modifications to the prior encoder.
By adopting the prior encoder to handle longer sound events while maintaining coherence and fidelity, we effectively tackled the challenges posed by the category-to-sound problem.
By adapting VITS to category-to-sound task, we propose Variational Inference for Foley sound Synthesis (VIFS). 
Our proposed method showcases the ability to generate high-quality sound clips with diversity, bridging the gap between category information and realistic audio representation for Foley sound synthesis. The official implementation code is available on GitHub\footnote{https://github.com/junjun3518/vifs}.

%% file: sections/2_VIFS.tex
\begin{figure}[t!]
  \centering
  \includegraphics[width=.95\linewidth]{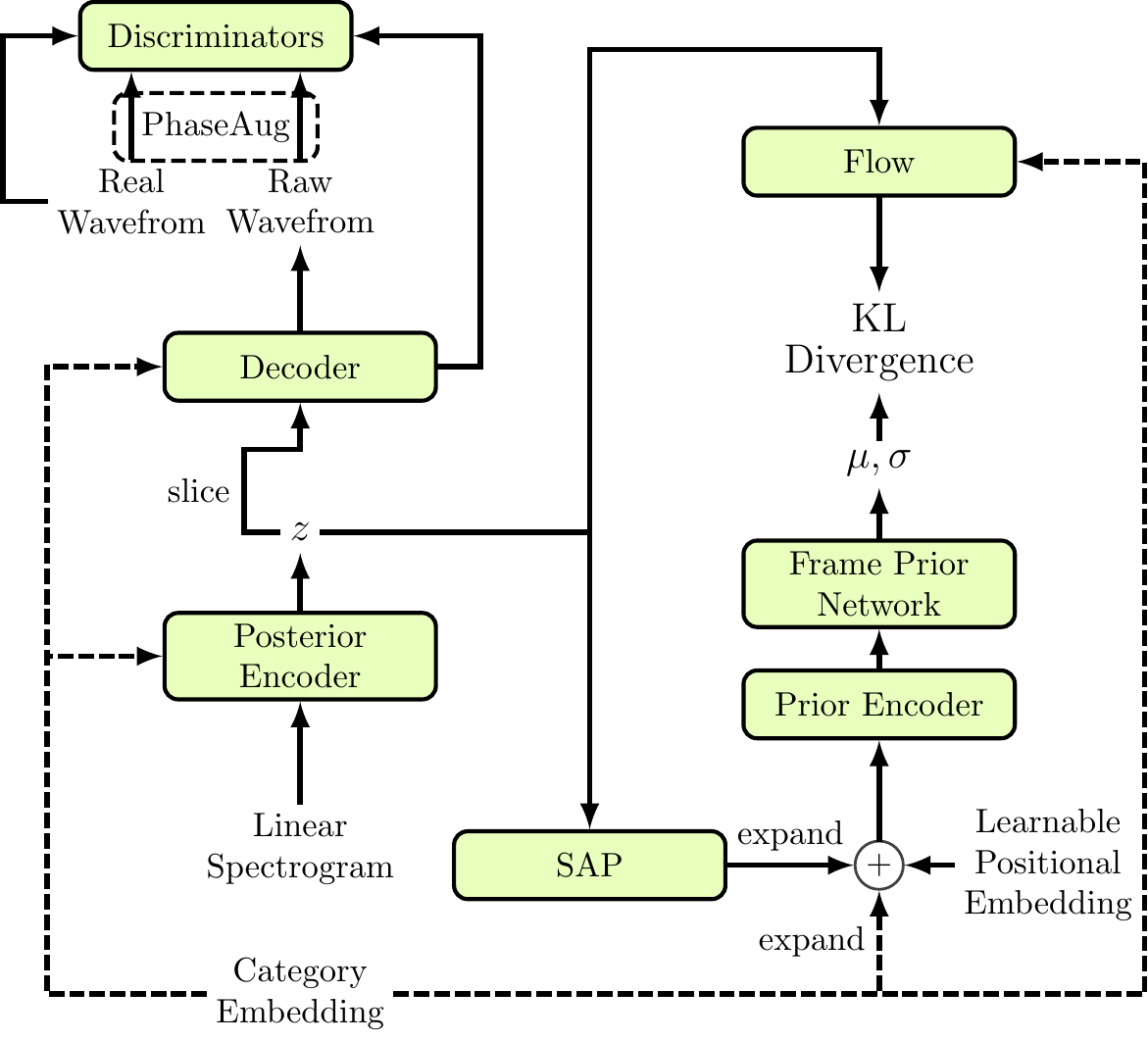}
  \vspace{-5pt}
  \caption{Overview of training procedure of VIFS.}
  \label{fig:model_figure}
\end{figure}

VIFS is heavily inspired by TTS studies. 
The architecture of VIFS is based on VITS \cite{vits}, therefore it consists of posterior encoder, prior encoder, flow, decoder, and discriminators.
Furthermore, we have applied various modifications by referring to previous works in speech synthesis.
Figure \ref{fig:model_figure} illustrates the overall training process. 

\subsection{Category Embedding}
VITS modules, including the posterior encoder, the flow, and the decoder, are conditioned by the speaker embedding to provide conditions for the specific speaker in the dataset.
Instead of the speaker embedding in VITS, we adopt category embedding for FSS to generate sounds for specific categories.
The dimensionality of the category embedding is identical to the hidden dimension of the coupling layers, which is 192.
Similar to the speaker embedding in VITS, the category embedding conditions the posterior encoder, the decoder, and the flow.
Furthermore, the category embedding serves as an input to the prior encoder, and its dimensionality is expanded to 344 to accommodate the time length of the latent variables from the prior encoder.
Considering that prior encoder in VITS takes phonemes as input, category embedding in VIFS replaces pronunciation information as well.

\subsection{Posterior Encoder and Flow}
We used the architecture of the posterior encoder and flow of VITS without modification.
However, FSS do not require monotonic alignment search (MAS), which is used for phoneme duration search in TTS.
Therefore, VIFS omitted MAS and just calculates Kullback–Leibler (KL) divergence without MAS and length regulator for prior latent variables. 

\subsection{Prior Encoder}\label{prior}
Different from TTS, the challenge requires to synthesize diverse sounds solely based on a single category index. 
Unlike phonemes, which typically have a duration of less than 100 million seconds, sound events span a much longer duration.
For DCASE 2023 Challenge Task 7, the sound events can reach a maximum length of 4 seconds.
To address this distinction and consider the different requirements of category-based sound synthesis, careful modifications need to be made to the prior encoder which encodes category index into various sound representations.

From the early experiments which did not consider clip-level consistency and were only conditioned by time-expanded category embedding,
the model generated sounds often containing different events and abrupt ends, compromising naturalness. 
For example, generated dog bark sound often involved barking from other dogs or even the echo of a gunshot, which corresponds to different category. Some sounds abruptly ended in the middle of events and restarted.
To address these problems, we introduce additional conditions for the prior encoder including category embedding, learnable positional embedding, and latent conditioning.

First, to generate sounds with clip-level consistency, the prior encoder needs to model latent variables considering temporal position. 
Thus, we add learnable positional embedding along time axis to expanded category embedding for conditioning positional information. 
The learnable positional embedding has a fixed dimension, which has an identical dimension with expanded category embedding.
In addition, we apply large-kernel frame prior networks \cite{visinger} to give more positional information to the model.
However, we observed that while position information improves synthesized sounds' consistency within each clip, it lowers the diversity within generated dataset.

Second, to enhance diversity, we tried adding time-expanded Gaussian noise to expanded category embedding.
However, just adding random Gaussian noise significantly increases KL divergence between flow outputs and prior latent variables which states given data is not modeled well with those structures. 
Therefore, we give another condition from the posterior latent variables to the prior encoder instead.
The posterior latent variables $z$ are compressed to a single vector by self-attentive pooling (SAP) \cite{sap} and expanded in time dimension to match the size of category embedding.
During the inference, Gaussian noise is sampled for input instead of SAP latent variables since we would not have posterior latent during inference.
To enforce the compressed vectors' distribution to standard normal Gaussian, we add L2 loss for the mean and standard deviation of the compressed vectors in a batch.

\subsection{GAN}
From the original GAN architecture of VITS which was adopted from HiFi-GAN \cite{hifigan}, we modified several features following Lee \etal \cite{pits}. 
This reflects the methodologies of Avocodo \cite{avocodo} which removes unintended artifacts such as aliasing and imaging artifacts.
The resultant GAN architecture is composed of the decoder and the discriminator in Figure \ref{fig:model_figure}. 
In addition, we also adopt PhaseAug \cite{phaseaug} for adversarial networks to prevent periodicity artifacts of the non-autoregressive vocoder structure in VIFS.

\subsection{Implementation Details}
Without the prior encoder, other details are identical to those of VITS \cite{vits}.
The frame prior network consists of 6 residual layers, each of which incorporates leaky ReLU using a negative slope 0.2, Conv1D with a kernel size of 35, and a residual connection. 
SAP calculates computes attention using the mean of the posterior latent variables.
4 seconds length of sounds are corresponding to spectrogram length 344 by the short-time Fourier transform with size 1024 and hop size 256.
This length serves as the expansion size for the category embedding and Gaussian input to the prior encoder.
The model training was performed using 4 V100 GPUs.

%% file: sections/3_experiments.tex
\subsection{Dataset}

We only use given dataset from the DCASE 2023 Challenge Task 7 \cite{fss}, which is sampled from UrbanSound8K \cite{urbansound}, FSD50K \cite{fsd50k}, and BBC Sound Effects\footnote{https://sound-effects.bbcrewind.co.uk/}.
This dataset consists of 7 distinct categories and total 4,850 files as illustrated in Table \ref{dataset}.
All files within the dataset are mono recordings with a bit depth of 16 bits, a sampling rate of 22,050 Hz, and a duration of 4 seconds.
\begin{table}[h]
    \centering
    \caption{Number of files on each category.}
    \label{dataset}
    \begin{tabular}{l|ll}
    \toprule
    Class ID & Category & Number of clips\\
    \midrule
    0&	DogBark&	617\\
    1&	Footstep&	703\\
    2&	GunShot&	777\\
    3&	Keyboard&	800\\
    4&	MovingMotorVehicle&	581\\
    5&	Rain&	741\\
    6&	Sneeze/Cough&	631\\  
    \bottomrule
    \end{tabular}
\end{table}

\subsection{Data Length}
\begin{figure}[t]
\centerline{\includegraphics[width=8.5cm]{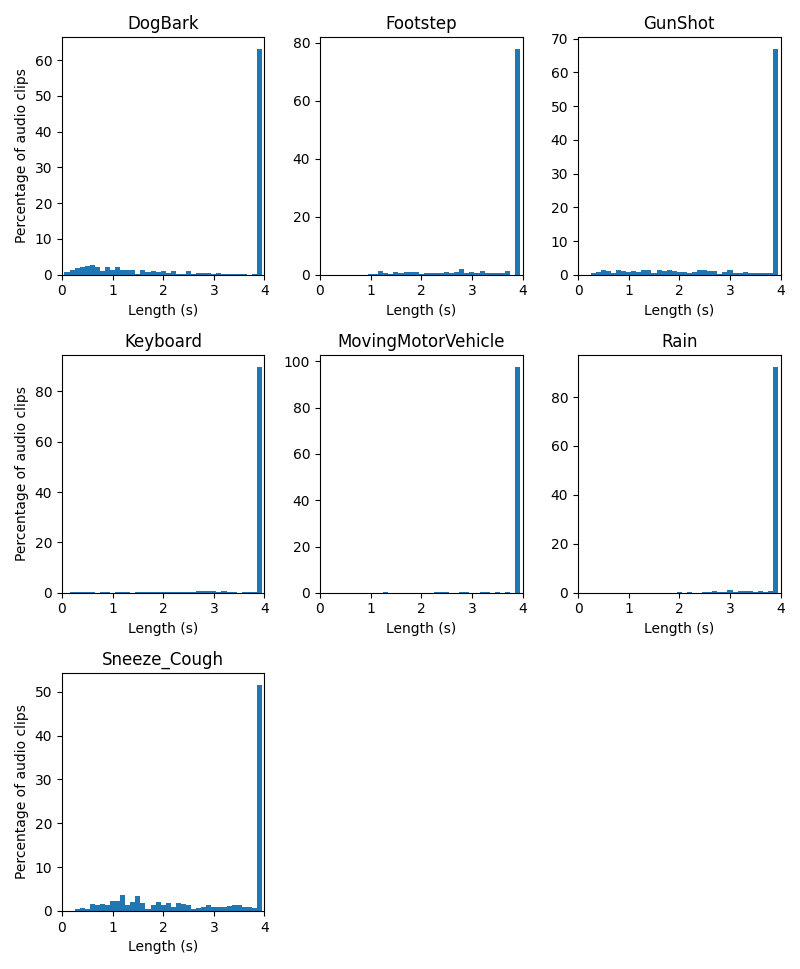}}
\vspace{-5pt}
\caption{Histograms of data lengths after removing zero padding.}
\label{fig:data_len}
\vspace{-15pt}
\end{figure}

The training dataset provided has been zero-padded to align all samples to a duration of 4 seconds \cite{fss}.
To consider the adverse effect of zero padding applied to the training dataset, we found the true data length by removing the zero padding and showed a histogram plots of each category on Figure \ref{fig:data_len}. Total number of audio clips corresponding to each category is shown in Table \ref{dataset} for the reference.
From the histograms, we can observe that more than half of the sound clips in training dataset are 4 seconds long before zero padding. 
These would be mostly audio clips trimmed at 4 seconds, though they were longer at first. From the histogram, we can observe that for categories of keyboard, moving motor vehicle and rain, the clips those are 4 seconds long are almost or more than 90\%. 
It is due to their characteristics that they usually occur longer than 4 seconds.
On the other hand, categories such as dog bark, gun shot and sneeze \& cough, the clips with 4 seconds long are less than 70\%. These sound events usually happen shortly. 
Nonetheless, total 76.1\% of training dataset is composed of sound clips 4 seconds long.
Also, among the files those are 4 second-long, there are noise from other sound sources at front and behind of the actual sound events corresponding to the category.
Therefore, we concluded that taking account of the zero padded would not significantly improve the model and we rather tried to exclude the effect of zeros padded or other noises within the sound clips within the prior encoder architecture as discussed in Section \ref{prior}.

\subsection{Evaluation Metric}
We used Fr\'echet Audio Distance (FAD) to evaluate traind model \cite{FAD}.
FAD is an object evaluation metric that measures the difference between distributions between the training dataset and generated dataset for each categories.
The distributions are composed of the representations of audio clips, extracted by inserting the audio clips to a VGGish model trained using AudioSet \cite{audioset}.
Since we perform FAD on the distributions of generated dataset and evaluation dataset, lower FAD implies that generated dataset is closer to evaluation dataset thus better the performance of FSS.
While FAD is the best option to perform objective evaluation on sound generative model, we should note the limitations of FAD: the trained VGGish model is not guaranteed to sufficiently working well on classification.
In addition, whether it is good classifier or not, the representation extracted from the VGGish model might not ``represent" the audio clips well too.
With these limitations, subjective tests are required to evaluate quality of generated models more thoroughly.

\subsection{Checkpoints and Noise Scales}\label{cans}
We used FAD to sort out the best models for each category, and chose models with four checkpoints and different noise scales. 
The noise scale is a factor applied to obtain the prior representation, and it is multiplied with Gaussian random vectors.
A higher noise scale introduces more variance in the generated sound clips, but it may also lead to the generation of clips that significantly differ from the samples in the training dataset. 
Conversely, a lower noise scale produces generated sound clips that are closer to the training dataset samples but with reduced variance.
To determine the optimal settings, we selected six checkpoints corresponding to 270k, 290k, 310k, 330k, 350k, and 370k steps. Among these checkpoints, we identified that four models achieve the best FAD for specific categories.
We then fine-tuned the noise scale for each model to further optimize the category-wise FAD. We first conducted tests using noise scales ranging from 0.25 to 1.5 with an interval of 0.25, followed by additional tests with noise scales from 0.5 to 1.0 with an interval of 0.1.

%% file: sections/4_results_and_discussion.tex
\subsection{Results}
\input{figures/result_table}

\begin{figure}[t]
\centerline{\includegraphics[width=8.5cm]{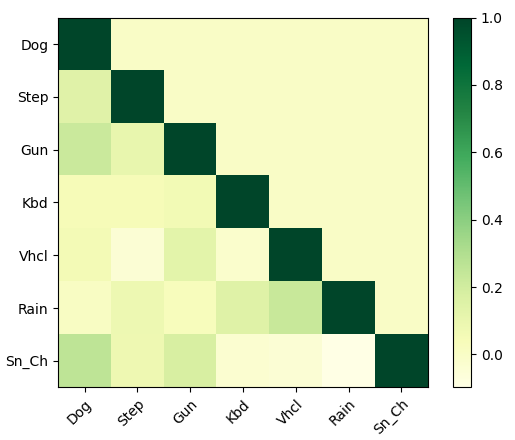}}
\vspace{-5pt}
\caption{Cosine similarity between trained category embedding. Upper triangular region is excluded due to the symmetry.}
\label{fig:embed}
\vspace{-15pt}
\end{figure}

Table \ref{tab:FAD} shows category-wise and average FAD on the baseline, six VIFS models and an ensemble VIFS model. The six VIFS models are chosen by procedure described on Section \ref{cans}, and each represents model checkpoints and noise scales corresponding to each category's best category-wise FAD, as shown in Table \ref{tab:FAD} on third to eighth columns.
The best FAD for each category are highlighted as bold. To further optimize the best model, we made an ensemble model by selecting the best model for each category, as indicated in the right column of Table \ref{tab:FAD}. For the challenge submission, we selected three individual models and the ensemble model, with their corresponding indices provided in Table \ref{tab:FAD} in the second row.


From the results, VIFS outperformed the baseline for the dog bark, footstep, gunshot, and rain categories. However, VIFS still requires improvement to surpass the baseline for the keyboard, moving motor vehicle, and sneeze \& cough categories. The FAD for the moving motor vehicle category was exceptionally high, causing the averaged FAD of VIFS to fall short of surpassing the baseline. 
However, when we considered the moving motor vehicle category as an outlier and evaluated the averaged FAD without it, we observed that the ensemble model outperformed the baseline by 13.2\%.

\subsection{Category Embeddings}
We present the cosine similarities between the trained category embeddings in Figure \ref{fig:embed}. While we discuss four checkpoints in this study, we observe that their category embeddings are nearly identical, so we only display the embeddings corresponding to the checkpoint at 270k steps. Considering the symmetry between upper and lower triangle, we exclude the upper triangle in the visualization for brevity. Additionally, to illustrate the contrast from the cosine similarity of 1, we include the diagonal elements. As depicted in the Figure \ref{fig:embed}, it is evident that the embeddings are weakly correlated with each other. The largest cosine similarity values are 0.261, 0.229, and 0.224, corresponding to the pairs of dog bark with sneeze \& cough, moving motor vehicle with rain, and dog bark with gun shot, respectively. These values are sufficiently small, indicating that the category embeddings have been effectively trained to differentiate between the different sound categories. 

An interesting observation to discuss is that relatively higher cosine similarity values follow the acoustic characteristics of the sound event categories. Dog bark, gun shot, and sneeze \& cough are all impulsive sounds, characterized by accentuated early parts of the sound waveform. They may occur once or in successive repeats. Similarly, moving motor vehicle and rain share similarities in terms of their stationary nature, where the spectral characteristics of these sounds rarely or slowly change over time \cite{FDY}. Furthermore, both categories exhibit spectral characteristics that span a wide frequency range. In the earlier stages of the checkpoints, these categories appear to be mixed with each other in the generated sound clips. For instance, a sample of generated sneeze \& cough sound contained gun shots and dog barking sounds during successive repeats. Similarly, a sample of moving motor vehicle sound resembled the sound of rain.

%% file: figures/result_table.tex
\begin{table*}[t!]
    \centering
    \caption{Category-wise FAD with chosen checkpoints and noise scales. A lower FAD value indicates a better alignment between the distribution of the generated audio clips and the real audio clips in each category.}
    \begin{tabular}{l|p{1.2cm}|p{1.2cm}p{1.2cm}p{1.2cm}p{1.2cm}p{1.2cm}p{1.2cm}|p{1.2cm}}
    \toprule
        \textbf{}       & \textbf{baseline}    & \multicolumn{6}{l}{\textbf{VIFS}}            & \textbf{ensemble} \\
        \midrule
        \# submission        & \phantom{00}-        & \phantom{00}-     & \phantom{00}3            & \phantom{00}-             & \phantom{00}-             & \phantom{00}2             & \phantom{00}1             & \phantom{00}4    \\
        \midrule
        \# step              & \phantom{00}-        & \phantom{0}270k  & \phantom{0}270k           & \phantom{0}270k           & \phantom{0}290k           & \phantom{0}310k           & \phantom{0}330k           & \phantom{00}-    \\
        noise scale          & \phantom{00}-        & \phantom{00}0.6  & \phantom{00}0.7           & \phantom{00}1.0           & \phantom{00}0.8           & \phantom{00}0.6           & \phantom{00}0.8           & \phantom{00}-    \\
        \midrule
        dog bark             & 13.411              & 12.009            & 12.184                    & 11.489                    & 11.227                    & 10.388                    & \phantom{0}\textbf{8.805} & \phantom{0}8.805 \\
        footstep             & \phantom{0}8.109    & \phantom{0}7.461  & \phantom{0}6.968          & \phantom{0}\textbf{6.638} & \phantom{0}6.889          & \phantom{0}7.373          & \phantom{0}7.290          & \phantom{0}6.638 \\
        gunshot              & \phantom{0}7.951    & \phantom{0}7.535  & \phantom{0}\textbf{7.233} & 12.440                    & \phantom{0}9.860          & \phantom{0}8.091          & \phantom{0}9.392          & \phantom{0}7.233 \\
        keyboard             & \phantom{0}5.230    & 10.359            & \phantom{0}9.191          & \phantom{0}7.643          & \phantom{0}7.634          & \phantom{0}9.699          & \phantom{0}\textbf{6.387} & \phantom{0}6.387 \\
        moving motor vehicle & 16.108              & \textbf{34.429}   & 34.880                    & 37.516                    & 39.905                    & 37.056                    & 37.818                    & 34.429           \\
        rain                 & 13.337              & \phantom{0}7.200  & \phantom{0}6.703          & \phantom{0}7.184          & \phantom{0}7.201          & \phantom{0}\textbf{6.636} & \phantom{0}7.899          & \phantom{0}6.636 \\
        sneeze \& cough      & \phantom{0}3.770    & \phantom{0}9.505  & \phantom{0}9.674          & \phantom{0}9.656          & \phantom{0}\textbf{9.283} & \phantom{0}9.744          & 11.916                    & \phantom{0}9.283 \\
        \midrule
        average w/o vehicle  & \phantom{0}8.635    & \phantom{0}9.007  & \phantom{0}8.659          & \phantom{0}9.175          & \phantom{0}8.682          & \phantom{0}8.655          & \phantom{0}8.615          & \phantom{0}7.497 \\
        average              & \phantom{0}9.702    & 12.638            & 12.405                    & 13.224                    & 13.142                    & 12.712                    & 12.787                    & 11.344           \\
        \bottomrule
    \end{tabular}
    \label{tab:FAD}
\end{table*}

%% file: sections/5_conclusion.tex
In this work, we propose VIFS, an end-to-end variational inference for FSS. 
With our heavily modified prior encoder, we could generate consistent sounds for each inference with high quality.
In addition, techniques from speech synthesis increase the perceptual quality of synthesized sounds without multiple stages of training.
As a result, we improved FAD of four categories, dog bark, footstep, gun shot and rain when compared to the baseline. While FAD for other categories are still behind the baseline, we would need to perform subjective test to compare the quality of generated sound.